\documentclass{article}
\usepackage{spconf,amsmath,graphicx,hyperref}
\usepackage{amsmath}         
\usepackage[flushleft]{threeparttable} 
\usepackage{cite}
\usepackage{booktabs, multirow}
\usepackage{enumitem} 
\usepackage{amssymb}

\usepackage{tikz}
\usetikzlibrary{arrows.meta,positioning,fit,calc}
\tikzset{
  block/.style={draw, rounded corners, very thick, align=center, inner sep=3pt, minimum height=8mm, fill=white},
  tinyblock/.style={draw, rounded corners, thick, inner sep=2pt, minimum height=4mm, fill=white},
  arrow/.style={-Latex, very thick}
}

\title{Joint Estimation of Piano Dynamics and Metrical Structure with a Multi-task Multi-Scale Network}
\name{Zhanhong He$^{1}$, Hanyu Meng$^{2}$, Defeng (David) Huang$^{1}$, Roberto Togneri$^{1}$}
\address{
    $^{1}$The University of Western Australia, Perth, Australia\\
    $^{2}$The University of New South Wales, Sydney, Australia
}
\begin{document}
\ninept
\maketitle
\begin{abstract}
Estimating piano dynamic from audio recordings is a fundamental challenge in computational music analysis. In this paper, we propose an efficient multi-task network that jointly predicts dynamic levels, change points, beats, and downbeats from a shared latent representation. These four targets form the metrical structure of dynamics in the music score. Inspired by recent vocal dynamic research, we use a multi-scale network as the backbone, which takes Bark-scale specific loudness as the input feature. Compared to log-Mel as input, this reduces model size from 14.7 M to 0.5 M, enabling long sequential input. We use a 60-second audio length in audio segmentation, which doubled the length of beat tracking commonly used. Evaluated on the public MazurkaBL dataset, our model achieves state-of-the-art results across all tasks. This work sets a new benchmark for piano dynamic estimation and delivers a powerful and compact tool, paving the way for large-scale, resource-efficient analysis of musical expression.

\end{abstract}
\begin{keywords}
Piano dynamics estimation, beat tracking, Bark-scale specific loudness, multi-scale network, multi-task learning
\end{keywords}
\section{Introduction}
\label{sec:intro}
The creation, comprehension and reproduction of music are fundamental aspects of human culture. In Western musical tradition, the term \textit{dynamics} refers to a coarse guide to the intended loudness. Indicated in a score by markings such as \textit{p} (\textit{piano}, soft) and \textit{f} (\textit{forte}, loud), dynamics are essential for shaping musical phrases, conveying emotion, and articulating structural differentiation \cite{cancinoChacon2018computational}. This vocabulary extends to a nuanced range of static levels, from \textit{pp} (\textit{pianissimo}) to \textit{ff} (\textit{fortissimo}), and includes gradual transitions like \textit{crescendo} and \textit{decrescendo}. The computational modeling of these expressive markings is valuable for music education and performance analysis \cite{narang2022analysis, zhang2025llaqo, park2025dynvisual}, as well as theory-informed music generation \cite{wu2024musiccontrolnet, he2025filling}.

The core challenge in estimating from audio lies in the inherent relativity and ambiguity of dynamic markings. A symbol like \textit{pianissimo} does not map to a fixed physical measurement like a decibel level. Instead, its interpretation is deeply contextual, influenced by musical style, performer's intent, and the acoustic environment \cite{kosta2014dynpiano, jones2023probingdyn}. This lack of a standardized ground truth has historically posed a significant challenge for machine learning algorithms, often leading to poor generalization among different performers or pieces \cite{kosta2016mapping, kosta2018mazurkabl}. 

To circumvent the ambiguity of music dynamics, a common strategy in music transcription and analysis is to adopt MIDI velocity as a proxy target \cite{van2014dynmidi, hawthorne2018onf, kim2023score, kim2024method}. This approach, however, introduces its own set of challenges. MIDI velocity reflects the performer's physical action rather than perceived loudness, and is confounded by the instrument's unique timbre and touch \cite{berndt2010modelling}. While automatic music transcription (AMT) systems can accurately estimate MIDI velocity from Yamaha Disklavier piano performances, generalizing this capability across diverse pianos remains unsolved \cite{edwards2024general}. This makes the subsequent task of regressing dynamic markings from the estimated MIDI velocity inherently unreliable.

Given these complexities, we propose an end-to-end multi-task learning approach to estimate piano dynamics and their underlying metrical structure directly from audio. Inspired by recent advances in vocal dynamics estimation \cite{narang2024automatic}, we use Bark-scale specific loudness (BSSL) as the input feature. The BSSL is processed by a multi-scale network backbone, adapted from \cite{li2023frame}, to extract a shared latent representation that reconciles the divergent temporal requirements of the distinct tasks. This unified latent representation is then channeled through a Multi-gate Mixture-of-Experts (MMoE) layer \cite{ma2018mmoe}, which generates specialized features for four task heads that jointly predict: (1) dynamic levels, (2) change points, (3) beat positions, and (4) downbeat positions. Together, these targets capture both the dynamic markings and their underlying metrical structure, since the beat and downbeat grid provides the rhythmic foundation of a musical score.

\begin{figure}[h]
\begin{center}
\includegraphics[width=\columnwidth]{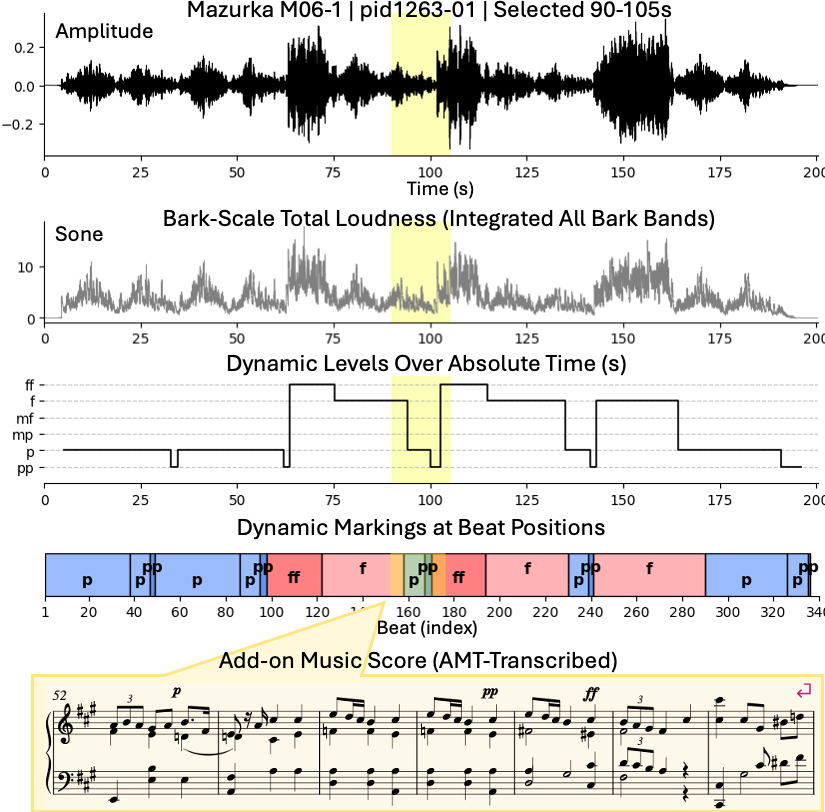}
\caption{Workflow for adding dynamic markings to an AMT-transcribed score from audio. The highlighted region (90–105 s of Chopin Mazurka Op. 6 No. 1) shows a case study. Dynamic markings are aligned with the musical beat grid, and a change point indicates a contextual shift in the dynamic levels.} 
\label{fg:bark}
\end{center}
\end{figure}

The primary application of our proposed multi-task framework is to enrich musical scores that possess reliable beat information but lack dynamic markings, a common scenario for music archives and the output of score-level AMT systems \cite{ijcai2024p862}. As depicted in Fig.~\ref{fg:bark}, this workflow aligns the model's predicted dynamics to a provided beat grid. Furthermore, the model's ability to jointly predict the beat and downbeat also allows it to operate fully end-to-end without a pre-existing music score, making it a versatile tool for large-scale piano performance analysis directly from audio.\footnote{Code and pre-trained models are available at: \url{https://github.com/zhanh-he/piano-dynamic-estimation}}


\section{Methodology}
\label{sec:method}

\subsection{Bark-Scale Specific Loudness}
\label{ssec:bark}
Bark-scale features are derived from a psychoacoustic model of human loudness perception across critical bands \cite{zwicker1999psychoacoustics}. While log-Mel spectrograms remain dominant in modern deep learning pipelines, the effectiveness of Bark-based representations is supported by extensive research. For instance, recent work has demonstrated that Bark-scale cepstral coefficients can improve speaker recognition under short-duration constraints \cite{zi2024bgcc}, and a high-resolution variant of BSSL has shown performance gains over log-Mel inputs for vocal dynamics estimation \cite{narang2024automatic}. Given the growing evidence supporting Bark-scale features, coupled with BSSL's established success in previous piano dynamics research \cite{kosta2016mapping}, we adopt it as the foundational feature for this work.

The primary challenge was the implementation of the BSSL extractor. Standardized toolboxes like MOSQITO \cite{san2021mosqito}, while compliant with ISO 532-1:2017 standard, are designed for robust sound quality assessment and thus mandate a 48 kHz sampling rate with constrained STFT parameters. Since most music and piano recordings are encoded at 44.1 kHz, using MOSQITO would necessitate upsampling, a computationally expensive process that can introduce interpolation artifacts. Therefore, we developed a custom BSSL feature extractor in the PyTorch framework, reproducing the \texttt{ma\_sone} function from the widely used Music Analysis MATLAB toolbox by Pampalk \cite{pampalk2004toolbox}, which is based on the algorithmic chain proposed in \cite{pampalk2002content}. This approach allows for the flexible parameter settings crucial for fair and efficient experimentation.

The feature extraction pipeline begins by converting the audio to mono, peak-normalizing it to $-1$\,dBFS. To ensure compatibility with a recent state-of-the-art (SOTA) beat tracking model \cite{foscarin2024beat}, the audio is resampled to 22.05\,kHz, and a short-time Fourier transform (STFT) is computed using a Hann window of length 1024 and a frame rate of 50\,fps (i.e., 20 ms per frame and hop size of 441). The resulting power spectrogram is then transformed into the BSSL through a series of psychoacoustic modeling steps. These include outer- and middle-ear weighting, grouping spectral energy into critical bands, modeling spectral masking, and nonlinear mapping to the perceptual \textit{sone} scale. This process yields the final input feature, a BSSL matrix $\mathbf{X}\in\mathbb{R}^{22 \times T}$, where $T$ denotes the number of time frames. The choice of 22 Bark bands is determined by the 11.025\,kHz Nyquist frequency, which fully covers the 22nd Bark band (extending to 9.5 kHz). While the total loudness provides an intuitive visualization (Fig.\ref{fg:bark}), the $22 \times T$ BSSL matrix is input to the model to preserve rich spectral details.

\begin{figure*}[t]
\begin{center}
\includegraphics[width=\textwidth]{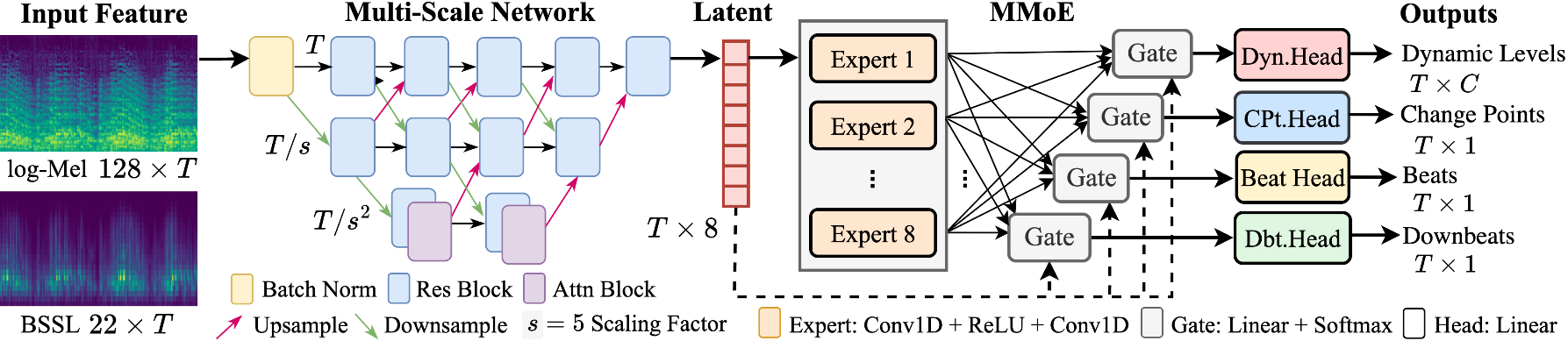}
\caption{Architecture of the proposed multi-task framework. A three-branch multi-scale network encodes either log-Mel or BSSL input. Branches operate at lengths $T$, $T/s$, and $T/s^{2}$, with outputs a latent sequence of shape $T\times8$. An 8-expert MMoE with four gates forms task-specific features, followed by linear heads to output four distinct targets. The scaling factor $s$ is a tunable hyperparameter.} 
\label{fg:model}
\end{center}
\end{figure*}

\subsection{Model Architecture}
\label{ssec:model}
The proposed model architecture is illustrated in Fig. \ref{fg:model}. With divergent acoustic requirements, dynamic estimation requires a large temporal receptive field \cite{narang2024automatic}, whereas beat and downbeat tracking require high temporal resolution to locate transient onsets \cite{bock2020deconstruct}. To address this need for varied receptive fields, we adapt the multi-scale network from \cite{li2023frame} as a shared encoder. Given an $F{\times}T$ time–frequency feature as input, the features are first normalized via 2D Batch Normalization (requiring a dimension transposition). The encoder's branches are then built from a series of residual and self-attention blocks detailed in \cite{li2023frame}. The encoder operates on three such parallel branches at different temporal resolutions, corresponding to lengths of $T$, $T/s$, and $T/s^{2}$, where down- and upsampling by the scaling factor $s$ are achieved using strides in max-pooling and transposed convolutions. Our architecture departs from the original design \cite{li2023frame} in two key ways: we treat the scaling factor $s$ as a configurable hyperparameter rather than a fixed value of 3, and the encoder outputs a shared latent sequence $\mathbf{Z}\in\mathbb{R}^{T \times 8}$ for later processing, rather than feeding into a single-task classifier.

To mitigate negative transfer between our acoustically diverse tasks, we develop a MMoE module \cite{ma2018mmoe} as a task-aware decoder that processes the shared representation. The module consists of two main components:
\begin{itemize}[noitemsep, topsep=1pt, leftmargin=*, labelsep=0.4em]
\item \textbf{Shared Experts}: The module contains a pool of 8 shared experts. Each expert in $i=1, \dots, 8$ is a lightweight temporal convolutional block. It comprises two sequential 1D convolutional layers, each with a kernel size of 3 and a length-preserving padding, separated by a ReLU activation. All experts process the shared latent sequence $\mathbf{Z}$ in parallel to produce a set of expert outputs.
\item \textbf{Task-Specific Gates}: For each of the four tasks $k$, a dedicated gating network $G_k(\cdot)$ acts as a dynamic router. Each gate is implemented as a simple linear layer that takes the latent feature $\mathbf{z}_t$ at each time step $t$ and outputs a softmax-normalized weight vector $\mathbf{w}_k(t) \in \mathbb{R}^8$. This vector determines how to weigh the contributions of the different experts for that specific task.
\end{itemize}
The final feature vector for task $k$ at a given time step $t$, denoted $\mathbf{y}_{k,t}$, is computed as the weighted sum of gates with all expert outputs:
\begin{equation}
\mathbf{y}_{k,t} = \sum_{i=1}^{8} \mathbf{w}_{k,i}(t) \cdot \mathbf{e}_{i,t} \,\,\, \text{where} \,\,\, \mathbf{w}_k(t) = \operatorname{Softmax}(G_k(\mathbf{z}_t))
\label{eq:mmoe_corrected}
\end{equation}
where $\mathbf{e}_{i,t}$ is the output of the $i$-th expert, and the weight vector $\mathbf{w}(t)$ is produced by the task-specific gate, whose $i$-th component $\mathbf{w}_{k,i}(t)$ represents the importance of expert $i$ for task $k$ at that time step. This computation is performed at each frame to form the complete task-specific feature $\mathbf{Y}_k = [\mathbf{y}_{k,1}, \dots, \mathbf{y}_{k,T}]^\top \in \mathbb{R}^{T \times 8}$. These four specialized representations are then mapped to the final logits by separate linear heads.

\subsection{Task-Specific Post-Processing}
\label{ssec:post}
The model’s raw output is a sequence of frame-wise probabilities (i.e., logits), which we convert into discrete musical events via tailored post-processing. For beat and downbeat detection, we adopt the procedure from \cite{foscarin2024beat}, identifying events by thresholding probabilities at 50\% and applying peak-picking within a 70\,ms neighborhood ($\pm3$ frames at 50\,fps). For dynamic markings, i.e., the dynamic level of each detected beat is determined by taking the \texttt{argmax} of the class probabilities at that specific time frame. Finally, change points are determined in a two-step process: we first identify all frames where the probabilities exceed a 75\% threshold, and then snap each of these candidates to the nearest detected beat to determine its location. This final beat-snapping step is performed because, while not a strict musical rule, the annotations (e.g., dynamic markings) in the MazurkaBL dataset are exclusively beat-aligned \cite{kosta2018mazurkabl}, and this approach maintains consistency with prior work \cite{kosta2017thesis}.

\subsection{Loss Function}
\label{ssec:loss}
Training the multi-task model involves a composite loss that combines specialized objectives for each task. Our multi-task loss, $\mathcal{L}_\mathrm{MTL}$, is defined as the sum of four task-specific losses:
\begin{equation}
\mathcal{L}_{\mathrm{MTL}}
= \mathcal{L}_{\mathrm{Dyn}}
+ \mathcal{L}_{\mathrm{CPt}}
+ \mathcal{L}_{\mathrm{Beat}}
+ \mathcal{L}_{\mathrm{Dbt}}
\end{equation}
where each component is defined as follows. For the binary targets (change points, beats, and downbeats), the loss terms $\mathcal{L}_{CPt}$, $\mathcal{L}_{Beat}$, and $\mathcal{L}_{Dbt}$ employ the shift-tolerant weighted binary cross-entropy proposed in \cite{foscarin2024beat}. This function addresses two key challenges: it counteracts the extreme sparsity of targets by weighting positive frames more heavily, and it accommodates annotation timing imprecision by incorporating a $\pm3$ frame tolerance window. For the multiclass target, dynamics, the loss term $\mathcal{L}_{Dyn}$ is a standard cross-entropy, masked by the ground-truth beat positions. This marking enforces a data-driven prior on the dataset's annotation style (i.e., dynamic markings occur only on beats), guiding the model to ignore spurious inter-beat fluctuations.

\section{Experiments}
\label{sec:exp}

\subsection{Dataset}
\label{ssec:data}
We use the MazurkaBL dataset \cite{kosta2018mazurkabl}, a score-aligned corpus of 2,098 solo piano recordings covering 44 Chopin Mazurkas. After excluding two mazurkas with irregular dynamics annotations (M06-4 and M63-2), 1,999 recordings are used. While similar datasets like the recently introduced EME33 \cite{hung2024eme33} exist, MazurkaBL is the largest publicly available resource for studying notated dynamics versus performed loudness from audio. Its provision of score-aligned beat times and verified expressive markings has led to its wide adoption in music analysis research. Designed for different goals, other large datasets like MAESTRO \cite{hawthorne2019maestro} use MIDI velocity as a proxy for dynamics, while others such as ASAP \cite{foscarin2020asap} provide score-aligned annotations but lack dynamic markings.
\subsection{Implementation Details}
\label{ssec:impl}
We use a 5-fold cross-validation protocol, stratified by 44 mazurkas, to train and evaluate our model. For each fold, the model is trained for 120 epochs, with the best-performing checkpoint selected based on the F1 score on its respective validation set. The training configuration includes the AdamW optimizer with a learning rate of 3e-4, a batch size of 10, and a fixed random seed of 86. 

For data augmentation, we slice audio into 60-second segments with 50\% overlap during training, while no overlap is used for evaluation. Both BSSL (22 Bark bands) and log-Mel (128 mel bins) features are extracted using the same STFT parameters as in \cite{foscarin2024beat} to ensure identical temporal resolution. The model employs a compact configuration with an empirically optimal temporal scaling factor of $s=5$. It predicts dynamic levels across $C=6$ classes: a \textit{blank} class for silence before the first annotation, and the five dynamic classes (\textit{pp, p, mf, f, ff}) from the MazurkaBL dataset. When run on a NVIDIA RTX 3090 (24 GiB), a full 5-fold run completed in approximately 20 hours, with peak GPU memory usage of 4 GiB.

\begin{table*}[t]
\centering
\caption{Performance comparison of the proposed model against all baselines. Per-task F1 scores (\%) are reported as mean $\pm$ standard deviations over 5-fold cross-validation. PELT is an algorithmic method with no trainable parameters, while the ANN did not report this attribute. The best score is highlighted in bold.}
\label{tab:main_results}
\begingroup
\fontsize{9pt}{11pt}\selectfont
\setlength{\tabcolsep}{6pt}\renewcommand{\arraystretch}{1.0}
\begin{tabular}{l|l|c|c|c|c|r}
\toprule
Method & Feature & Dynamic F1 & Change Point F1 & Beat F1 & Downbeat F1 & \# Params \\
\midrule
ANN \cite{kosta2017thesis}    & BSSL & 29.4 & -- & -- & -- & n/a \\
PELT  \cite{kosta2017thesis}  & BSSL & -- & $10.8$  & -- & -- & n/a \\
TCN+DBN \cite{bock2020deconstruct}  & log-Mel & -- & -- & $60.9 \pm 1.8$ & $30.4 \pm 1.3$ & 0.1 M \\
Beat This \cite{foscarin2024beat}   & log-Mel         & -- & -- & $80.5 \pm 2.7$ & $52.8 \pm 6.2$ & 20.3 M \\
\midrule
Single-task Multi-scale Network     & BSSL & 50.6 $\pm$ 10.1 & 21.0 $\pm$ 9.9 & 84.0 $\pm$ 1.5 & 45.0 $\pm$ 1.7 & 0.4 M \\
\quad w/o. BSSL & log-Mel         & 50.4 $\pm$ 11.1 & 17.5 $\pm$ 5.4 & 83.8 $\pm$ 1.8 & 54.7 $\pm$ 7.5 & 13.3 M \\
\midrule
Multi-task Multi-scale Network (Proposed)  & BSSL & \textbf{54.4 $\pm$ 8.9} & \textbf{26.1 $\pm$ 9.7} & \textbf{84.1 $\pm$ 1.3} & 55.2 $\pm$ 4.2 & 0.5 M \\
\quad w/o. BSSL & log-Mel      & 50.8 $\pm$ 10.9 & 23.1 $\pm$ 6.1 & 83.7 $\pm$ 1.7 & \textbf{58.5 $\pm$ 6.2} & 14.7 M \\
\bottomrule
\end{tabular}
\endgroup
\end{table*}

\subsection{Baselines}
\label{ssec:baselines}
We compare the proposed multi-task network against the single-task network and task-specific baselines.

\noindent\textbf{Single-task Multi-scale Network.} We establish single-task baselines by adapting the multi-scale network from \cite{li2023frame}, integrating their model architecture (code publicly available) into our data handling pipeline and training an independent model for each target. The loss for each single-task network is the corresponding single term from our multi-task loss, with an empirically optimal temporal scaling factor of 5.

\noindent\textbf{Dynamic and Change Point Baselines.} We report published results from representative methods in \cite{kosta2017thesis}. This includes the Artificial Neural Networks (ANN) for dynamics, and the Pruned Exact Linear Time (PELT) algorithms for change points. These methods are tuning-intensive and non-end-to-end, so we report the literature scores rather than re-implement them.

\noindent\textbf{Beat and Downbeat Baselines.} We include the time-convolutional network (TCN) with a dynamic Bayesian network (DBN) post-processor \cite{bock2020deconstruct}, and the recent state-of-the-art transformer model, Beat This \cite{foscarin2024beat}. Both models can estimate beats and downbeats simultaneously. We retrain them from scratch on the MazurkaBL dataset using their publicly available code and the same 5-fold protocol as ours.

\subsection{Evaluation Metrics}
\label{ssec:metrics}
Performance on all four tasks is evaluated using the F1 score. For beat and downbeat tracking, we report F1 with a \mbox{$\pm70$\,ms} tolerance, consistent with prior work \cite{bock2020deconstruct,foscarin2024beat}. Evaluation of both dynamics and change points is consistent with \cite{kosta2017thesis}. For dynamics, the model's continuous output (dynamic level curve) is first sampled at each ground-truth beat location, and these values are then discretized into the corresponding dynamic markings. This converts the frame-wise prediction into a sequence of beat-wise labels, which are evaluated using a macro-averaged F1 score across five dynamic classes (\textit{pp, p, mf, f, ff}, excluding the \textit{blank} class). For change points, their predictions are snapped to the nearest ground-truth beat before being evaluated with a standard F1 score. This beat-wise alignment for both tasks turns the evaluation into a direct, musically metrical index-based comparison, thus requiring no additional timing tolerance.




\section{Results}
\label{sec:results}
\subsection{Main Result}
As presented in Table~\ref{tab:main_results}, our proposed multi-task model achieves SOTA performance in dynamics and change point estimation, while performing competitively on the remaining tasks. The effectiveness of the multi-task learning paradigm is underscored by the model's superior performance relative to its single-task counterpart using the same BSSL features. This includes significant F1 score improvements in dynamics (+3.8\%), change points (+5.1\%), beats (+0.1\%), and downbeats (+10.2\%). Beyond these quantitative improvements, the multi-task model offers considerable practical utility. It operates within a highly parameter-efficient framework (0.5 M vs. 4$\times$ single-task 0.4 M) and can utilize its own predicted beat positions for post-processing, enabling practical application on unannotated audio.

Our analysis also reveals a strong task-feature dependency: BSSL features are optimal for dynamics, change points, and beats estimation, whereas log-Mel features are preferable for downbeat tracking. A primary advantage of BSSL is its compactness (22 Bark bins vs. 128 Mel bins in our STFT setup). Within our multi-task multi-scale network, which relies on convolutional residual blocks, this smaller input dimension can reduce the model's trainable parameters from 14.7 M to just 0.5 M. This significantly smaller footprint enables the model to process longer audio sequences, directly benefiting tasks that require long-term temporal information and highlighting BSSL's potential for a wider range of musical applications with long-term dependencies.

\subsection{Ablation Study}
To validate our design choices, we conduct a comprehensive ablation study, introducing an average score (calculated by averaging the 5-fold mean F1 scores from the four tasks) to measure global performance. We systematically evaluate four configurations against our full model: (i) removing the MMoE module; (ii) disabling the multi-scale functionality by setting the scaling factor $s=1$; (iii) removing data augmentation by using non-overlapping 60-second audio segments in training stage; and (iv) reducing the input audio length from our default setting 60-second to 30-second (same length as in \cite{foscarin2024beat}). As detailed in Table \ref{tab:ablation}, each of the proposed components and training choices contributes meaningfully to the model's final performance, with the extended 60-second input context providing a significant advantage in dynamics-related tasks.

\begin{table}[t]
\centering
\caption{Ablation study of the multitask network with BSSL features. Per-task F1 scores (\%, mean only) and their average are reported over 5-fold, showing impacts of disabling key components.}
\label{tab:ablation}
\begingroup
\fontsize{9pt}{11pt}\selectfont
\setlength{\tabcolsep}{3.3pt}\renewcommand{\arraystretch}{1.0}
\begin{tabular}{l|c|c|c|c|c}
\toprule
Setting & Dyn F1 & CPt F1 & Bt F1 & Dbt F1 & Average \\
\midrule
Proposed         & \textbf{54.4} & \textbf{26.1} & \textbf{84.1} & \textbf{55.2} & \textbf{55.0}   \\
\,\,\, w/o. MMoE        & 52.8 & 22.0 & 82.9 & 51.8 & 52.4 \\
\,\,\, w/o. Temp. Scal. & 50.5 & 13.3 & 80.3 & 41.9 & 46.5 \\
\,\,\, w/o. Data Augm.  & 50.5 & 19.6 & 83.2 & 51.7 & 51.2 \\
\,\,\, uses 30s Segment & 49.1 & 19.2 & 83.4 & 52.7 & 51.1 \\
\bottomrule
\end{tabular}
\endgroup
\end{table}

\section{Conclusion}
\label{sec:concl}
In this paper, we proposed a compact multi-task, multi-scale network that jointly estimates piano dynamics, change points, beats, and downbeats directly from audio. Using Bark-scale specific loudness as input and an MMoE decoder, our model leverages a 60-second temporal context while remaining highly parameter-efficient with only 0.5 M parameters. Evaluated on the MazurkaBL dataset, our model achieves state-of-the-art results for dynamics and change point detection, while demonstrating competitive performance in beat and downbeat tracking. This demonstrates not only the model's practical utility but also its significant potential for broader applications. Future work will focus on combining our proposed model with score-level piano transcription systems. Such an end-to-end pipeline could produce music scores with dynamic markings from the performance audio, but developing appropriate evaluation methods for such comprehensive outputs presents a new challenge.

\bibliographystyle{IEEEbib}
\bibliography{dynest}

@inproceedings{foscarin2024beat,
  author={Foscarin, Francesco and Schlüter, Jan and Widmer, Gerhard},
  title={Beat This! Accurate Beat Tracking Without DBN Postprocessing},
  booktitle={Proc. ISMIR},
  year={2024},
  pages={962--969},
  nk_doi={10.5281/zenodo.14877491},
  bk_url={https://arxiv.org/pdf/2407.21658},
  comment={提到了现有的beat estimation模型, 当前的beat est的SOTA}
}

@inproceedings{bock2020deconstruct,
  author={Böck,Sebastian and Davies,Matthew E.P.},
  title={Deconstruct, Analyse, Reconstruct: How to Improve Tempo, Beat, and Downbeat Estimation},
  booktitle={Proc. ISMIR},
  year={2020},
  pages={574--582},
  bk_url={https://zenodo.org/records/4245498},
  bk_code={https://colab.research.google.com/drive/1tuOqNyO9gdMmYJsj33fP_QOfpRsm2tmt?usp=sharing#scrollTo=qz8UkWjNJDsx},
  bk_github={https://tempobeatdownbeat.github.io/tutorial/intro.html},
  comment={TCN beat estimation模型}
}

@book{zwicker1999psychoacoustics,
  title        = {Psychoacoustics: Facts and Models},
  author       = {Zwicker, Eberhard and Fastl, Hugo},
  series       = {Springer Series in Information Sciences},
  volume       = {22},
  edition      = {2nd, updated},
  publisher    = {Springer},
  address      = {Berlin},
  year         = {1999},
  doi          = {10.1007/978-3-662-11962-1}
}

@inproceedings{pampalk2002content,
  author={Pampalk,Elias and Rauber,Andreas and Merkl,Dieter},
  title={Content-based Organization and Visualization of Music Archives},
  booktitle={Proc. ACM Int. Conf. Multimedia},
  year={2002},
  pages={570--579},
  bk_publisher={ACM},
  isbn={158113620X},
  bk_pdf={https://ofai.at/papers/oefai-tr-2002-30.pdf},
  doi={10.1145/641007.641121},
  comment={文字描述怎么实现bark_sone}
}

@inproceedings{pampalk2004toolbox,
  author={Pampalk,Elias},
  title={A Matlab Toolbox to Compute Music Similarity from Audio},
  booktitle={Proc. ISMIR},
  year={2004},
  pages={254--257},
  bk_code={https://www.pampalk.at/ma/documentation.html},
  bk_ppt={https://www.ofai.at/~elias.pampalk/publications/pam_ismir04_poster.pdf},
  comment={这篇文章ma_sone 2007年的matlab toolbox文章，但我觉得这个不准确}
}

@inproceedings{berndt2010modelling,
  author={Berndt,Axel and Hähnel,Tilo},
  title={Modelling Musical Dynamics},
  booktitle={Proceedings of the 5th Audio Mostly Conference: A Conference on Interaction with Sound},
  year={2010},
  pages={1--8},
  bk_doi={10.1145/1859799.1859817},
  bk_url={https://dl.acm.org/doi/pdf/10.1145/1859799.1859817},
  comment={这篇文章是关于MIDI velocity -- dynamic marking 转换的基准方法，可以引用并衔接到HPT之类的模型后面}
}

@inproceedings{van2014dynmidi,
  author={van Herwaarden,Sam and Grachten,Maarten and de Haas,W. Bas},
  title={Predicting Expressive Dynamics in Piano Performances using Neural Networks},
  booktitle={Proc. ISMIR},
  year={2014},
  pages={45--52},
  bk_address={Taipei, Taiwan},
  bk_doi={10.5281/zenodo.1416678},
  bk_url={https://phenicx.upf.edu/system/files/publications/nn_piano.pdf},
  comment={Machine Learning (RBM)方法。MIDI velocity到dynamic marking转换的工作，指出这并不容易，吃很多环境因素和声学考量。小心这个reference可能有名字出错}
}

@inproceedings{kosta2014dynpiano,
  author={Kosta,Katerina and Bandtlow,Oscar F. and Chew,Elaine},
  title={Practical Implications of Dynamic Markings in the Score: Is Piano Always Piano?},
  booktitle={AES 53rd International Conference: Semantic Audio},
  year={2014},
  pages={1--3},
  bk_address={London, UK},
  url={https://secure.aes.org/forum/pubs/conferences/?elib=17120}
}

@phdthesis{kosta2017thesis,
  author={Kosta,Katerina},
  title={Computational Modelling and Quantitative Analysis of Dynamics in Performed Music},
  school={Centre for Digital Music, Queen Mary University of London},
  year={2017},
  bk_url={https://qmro.qmul.ac.uk/xmlui/handle/123456789/30712},
  comment={在77面的表格6.2中提及了change_point的分数}
}

@inproceedings{kosta2018mazurkabl,
  author={Kosta,Katerina and Bandtlow,Oscar F. and Chew,Elaine},
  title={{MazurkaBL}: score-aligned Loudness, Beat, and Expressive Markings Data for 2000 Chopin Mazurka Recordings},
  booktitle={Proc. Int. Conf. Tech. for Music Notation and Rep.},
  year={2018},
  pages={85--94},
  bk_address={Montréal, Canada},
  bk_doi={10.5281/zenodo.1289691},
  bk_url={https://www.tenor-conference.org/proceedings/2018/12_Kosta_tenor18.pdf},
  comment={mazurka数据集的文章,讨论了这个数据集的几个应用,给出了几个论点:machine learning只能根据歌曲建模,不能泛化到同个作曲家,更不论是更广泛的乐曲中。}
}

@inproceedings{san2021mosqito,
  author={San Millán‑Castillo,Roberto and Latorre‑Iglesias,Eduardo and Glesser,Martin and Wanty,Salomé and Jiménez‑Caminero,Daniel and Álvarez‑Jimeno,José María},
  title={{MOSQITO}: an open‑source and free toolbox for sound quality metrics in the industry and education},
  booktitle={INTER-NOISE and NOISE-CON Congress and Conference Proceedings},
  year={2021},
  pages={1164--1175}
}

@inproceedings{narang2022analysis,
  author={Narang,Jyoti and Miron,Marius and Srinivasamurthy,Ajay and Serra,Xavier},
  title={Analysis of Musical Dynamics in Vocal Performances Using Loudness Measures},
  booktitle={Proc. Int. Conf. Digital Audio Effects (DAFx)},
  year={2022},
  pages={33--39},
  address={Vienna, Austria},
  bk_url={https://www.dafx.de/paper-archive/2022/papers/DAFx20in22_paper_8.pdf},
  comment={貌似引入了一种衡量学生和老师之间dyn相似度的评估指标 "Global Dynamic Similarity"}
}

@article{jones2023probingdyn,
  author={Jones,Gabriel and Friberg,Anders},
  title={Probing the Underlying Principles of Dynamics in Piano Performances Using a Modelling Approach},
  journal={Frontiers in Psychology},
  year={2023},
  volume={14},
  pages={1269715},
  doi={10.3389/fpsyg.2023.1269715},
  url={https://www.frontiersin.org/articles/10.3389/fpsyg.2023.1269715},
  comment={用机械学习方法从MIDI预测dyanmic markings，基于yamaha disklavier piano}
}

@article{zi2024bgcc,
  title={Improving Short-Duration Speaker Recognition by Joint Bark-Wavelet Acoustic Feature Coupling and Triplet Dual-Attention Mechanism Network},
  author={Yunfei Zi and Shengwu Xiong},
  journal={Wireless Personal Communications},
  volume={135},
  pages={1725--1746},
  year={2024},
  publisher={Springer}
}

@article{edwards2024general,
  author={Edwards, Drew and Dixon, Simon and Benetos, Emmanouil and Maezawa, Akira and Kusaka, Yuta},
  title={A Data-Driven Analysis of Robust Automatic Piano Transcription},
  journal={IEEE Signal Processing Letters},
  year={2024},
  volume={31},
  pages={681--685},
  bk_doi={10.1109/LSP.2024.3363646},
  comment={研究钢琴转录模型的泛化能力，提出现有模型对不同钢琴泛化能力不足，因为它们都是基于Yamaha Disklavier钢琴训练的}
}

@article{park2025dynvisual,
  author={Park,Eun Ji},
  title={Music Dynamics Visualization for Music Practice and Education},
  journal={Multimedia Tools and Applications},
  volume={84},
  pages={36145--36161},
  year={2025},
  doi={10.1007/s11042-025-20637-0},
  url={https://link.springer.com/article/10.1007/s11042-025-20637-0},
  mycomment={dynamic在教育中的应用}
}

@article{kosta2016mapping,
  author={Kosta,Katerina and Ramirez,Rafael and Bandtlow,Oscar F. and Chew,Elaine},
  title={Mapping Between Dynamic Markings and Performed Loudness: A Machine Learning Approach},
  journal={Journal of Mathematics and Music},
  volume={10},
  number={2},
  pages={149--172},
  year={2016},
  bk_doi={10.1080/17459737.2016.1177404},
  bk_url={https://repositori-api.upf.edu/api/core/bitstreams/91dc82be-2f30-4ae4-92ce-0851200db44d/content},
  comment={这篇文章是关于dynamic estimation的Machine Learning方法 利用了F1和CCI%作为评估 并提到跨performer(painist)的泛化对ML模型是不可能的}
}

@inproceedings{ma2018mmoe,
  author={Ma,Jiaqi and Zhao,Zhe and Yi,Xinyang and Chen,Jilin and Hong,Lichan and Chi,Ed H.},
  title={Modeling Task Relationships in Multi-task Learning with Multi-gate Mixture-of-Experts},
  booktitle={Proc. ACM SIGKDD Int. Conf. Knowl. Discov. Data Min.},
  year={2018},
  pages={1930--1939},
  doi={10.1145/3219819.3220007}
}

@inproceedings{li2023frame,
  author={Li,Dichucheng and Che,Mingjin and Meng,Wenwu and Wu,Yulun and Yu,Yi and Xia,Fan and Li,Wei},
  title={Frame-Level Multi-Label Playing Technique Detection Using Multi-Scale Network and Self-Attention Mechanism},
  booktitle={Proc. ICASSP},
  year={2023},
  pages={1--5},
  bk_url={https://arxiv.org/abs/2303.13272},
  comment={这篇文章是jyoti narang采用的CNN模型架构}
}

@inproceedings{narang2024automatic,
  author={Narang,Jyoti and Tamer,Nazif Can and De La Vega,Viviana and Serra,Xavier},
  title={Automatic Estimation of Singing Voice Musical Dynamics},
  booktitle={Proc. ISMIR},
  year={2024},
  pages={256--263},
  bk_doi={10.5281/zenodo.14877323},
  bk_url={https://arxiv.org/abs/2410.20540},
  comment={这篇文章是vocal dynamic estimation采用了CRNN模型架构}
}

@inproceedings{hawthorne2018onf,
  author={Hawthorne,Curtis and Stasyuk,Andriy and Roberts,Adam and Simon,Ian and Huang,Sageev and Dieleman,Sander and Elsen,Erich and Engel,Jesse and Eck,Douglas},
  title={Onsets and Frames: Dual-Objective Piano Transcription},
  booktitle={Proc. ISMIR},
  year={2018},
  pages={50--57},
  comment={how to convert MIDI to matrix representation}
}

@inproceedings{hawthorne2019maestro,
  author    = {Hawthorne, Curtis and others},
  title     = {Enabling Factorized Piano Music Modeling and Generation with the {MAESTRO} Dataset},
  booktitle = {Proc. ICLR},
  year      = {2019},
  bk_url    = {https://openreview.net/forum?id=r1lYRjC9F7},
  comment   = {MAESTRO数据集的文章, 记得去overleaf上确认下正确引用格式}
}

@inproceedings{foscarin2020asap,
  author={Foscarin,Francesco and McLeod,Andrew and Rigaux,Philippe and Jacquemard,Florent and Sakai,Masahiko},
  title={{ASAP}: a dataset of aligned scores and performances for piano transcription},
  booktitle={Proc. ISMIR},
  year={2020},
  pages={534--541}
}

@inproceedings{kim2023score,
  author={Kim, Hyon and Miron,Marius and Serra,Xavier},
  title={Score-Informed MIDI Velocity Estimation for Piano Performance by FiLM Conditioning},
  booktitle={Proc. of the Sound and Music Computing Conf. (SMC)},
  year={2023},
}

@inproceedings{kim2024method,
  author={Kim, Hyon and Serra,Xavier},
  title={A Method for MIDI Velocity Estimation for Piano Performance by a U-Net with Attention and FiLM},
  booktitle={Proc. ISMIR},
  year={2024},
  pages={304--310},
  bk_address={San Francisco, USA}
}

@inproceedings{hung2024eme33,
  author={Hung,Tzu-Ching and Tang,Jingjing and Armstrong,Kit and Lin,Yi-Cheng and Liu,Yi-Wen},
  title={EME33: A Dataset of Classical Piano Performances Guided by Expressive Markings with Application in Music Rendering},
  booktitle={Proc. of IEEE Int. Conf. on Big Data},
  year={2024},
  pages={3174--3180},
  doi={10.1109/BigData62323.2024.10826039},
}

@inproceedings{he2025filling,
  title={Filling MIDI Velocity using U-Net Image Colorizer},
  author={He, Zhanhong and Cooper, David and Huang, Defeng and Togneri, Roberto},
  booktitle={Proc. 17th Int. Symp. Computer Music Multidisciplinary Research},
  pages={949--960},
  year={2025},
}

@article{cancinoChacon2018computational,
  title={Computational models of expressive music performance: A comprehensive and critical review},
  author={Cancino-Chac{\'o}n, Carlos E and Grachten, Maarten and Goebl, Werner and Widmer, Gerhard},
  journal={Frontiers in Digital Humanities},
  volume={5},
  pages={25},
  year={2018},
  publisher={Frontiers Media SA},
  url={https://www.frontiersin.org/articles/10.3389/fdigh.2018.00025},
}

@article{wu2024musiccontrolnet,
  author={Wu,Shih‑Lun and Donahue,Chris and Watanabe,Shinji and Bryan,Nicholas J.},
  title={Music ControlNet: Multiple Time‑Varying Controls for Music Generation},
  journal={IEEE/ACM Trans. Audio, Speech, Lang. Process.},
  year={2024},
  volume={32},
  pages={2692--2703},
  doi={10.1109/TASLP.2024.3399026},
  url={https://doi.org/10.1109/TASLP.2024.3399026}
}

@inproceedings{zhang2025llaqo,
  author={Zhang,Huan and Cheung,Vincent K.M. and Nishioka,Hayato and Dixon,Simon and Furuya,Shinichi},
  title={LLaQo: Towards a Query-Based Coach in Expressive Music Performance Assessment},
  booktitle={Proc. ICASSP},
  year={2025},
  pages={1--5},
  doi={10.1109/ICASSP49660.2025.10890522},
}

@inproceedings{ijcai2024p862,
  title     = {End-to-End Real-World Polyphonic Piano Audio-to-Score Transcription with Hierarchical Decoding},
  author    = {Zeng, Wei and He, Xian and Wang, Ye},
  booktitle = {Proc. IJCAI},
  bk_publisher = {International Joint Conferences on Artificial Intelligence Organization},
  bk_editor    = {Kate Larson},
  pages     = {7788--7795},
  year      = {2024},
  bk_month     = {8},
  bk_note      = {AI, Arts \& Creativity},
  bk_doi       = {10.24963/ijcai.2024/862},
  bk_url       = {https://doi.org/10.24963/ijcai.2024/862},
}

\end{document}